\documentclass[aps,prd,twocolumn,groupedaddress,amssymb,eqsecnum,showpacs,epsfig]{revtex4}
\usepackage{graphicx}
\usepackage{bm}
\usepackage{dcolumn}
\usepackage{amsmath}
\usepackage{amssymb}
\def \lleq {\lower0.9ex\hbox{ $\buildrel < \over \sim$} ~}
\def \ggeq {\lower0.9ex\hbox{ $\buildrel > \over \sim$} ~}

\def \beq  {\begin{equation}}
\def \eeq  {\end{equation}}
\def \ber  {\begin{eqnarray}}
\def \eer  {\end{eqnarray}}

\def\apj{{Astroph.\@ J.\ }}

\def\aj{{Astron.\@ J.\ }}

\begin{document}
\newcommand{\newc}{\newcommand}

\newc{\be}{\begin{equation}}
\newc{\ee}{\end{equation}}
\newc{\ba}{\begin{eqnarray}}
\newc{\ea}{\end{eqnarray}}
\newc{\bea}{\begin{eqnarray*}}
\newc{\eea}{\end{eqnarray*}}
\newc{\D}{\partial}
\newc{\ie}{{\it i.e.} }
\newc{\eg}{{\it e.g.} }
\newc{\etc}{{\it etc.} }
\newc{\etal}{{\it et al.}}
\newcommand{\nn}{\nonumber}
\newc{\ra}{\rightarrow}
\newc{\lra}{\leftrightarrow}
\newc{\lsim}{\buildrel{<}\over{\sim}}
\newc{\gsim}{\buildrel{>}\over{\sim}}
\title{Comparison of the Legacy and Gold SnIa Dataset Constraints on Dark Energy Models}
\author{S. Nesseris$^a$ and L. Perivolaropoulos$^b$ }
\affiliation{Department of Physics, University of Ioannina, Greece
\\ $^a$ e-mail: me01629@cc.uoi.gr, $^b$ e-mail:
leandros@cc.uoi.gr}
\date{\today}

\begin{abstract}
We have performed a comparative analysis of three recent and
reliable SnIa datasets available in the literature: the Full Gold
(FG) dataset (157 data points $0<z<1.75$), a Truncated Gold (TG)
dataset (140 data points $0<z<1$) and the most recent Supernova
Legacy Survey (SNLS) dataset (115 data points $0<z<1$). We have
analyzed and compared the likelihood of cosmological constant and
dynamical dark energy parametrizations allowing for crossing of
the phantom divide line (PDL). We find that even though the
constraints obtained using the three datasets are consistent with
each other at the 95\% confidence level, the latest (SNLS) dataset
shows distinct trends which are not shared by the Gold datasets.
We find that the best fit dynamical $w(z)$ obtained from the SNLS
dataset does not cross the PDL $w=-1$ and remains above and close
to the $w=-1$ line for the whole redshift range $0<z<1$ showing no
evidence for phantom behavior. The LCDM parameter values
($w_0=-1$, $w_1=0$) almost coincide with the best fit parameters
of the dynamical $w(z)$ parametrizations. In contrast, the best
fit dynamical $w(z)$ obtained from the Gold datasets (FG and TG)
clearly crosses the PDL and departs significantly from the PDL
$w=-1$ line while the LCDM parameter values are about $2\sigma$
away from the best fit $w(z)$. In addition, the
$(\Omega_{0m},\Omega_\Lambda)$ parameters in a LCDM
parametrization without a flat prior, fit by the SNLS dataset,
favor the minimal flat LCDM concordance model. The corresponding
fit with the Gold datasets mildly favors a closed universe and the
flat LCDM parameter values are $1\sigma - 2\sigma$ away from the
best fit $(\Omega_{0m},\Omega_\Lambda)$.
\end{abstract}
\maketitle

\section{Introduction}

Current cosmological observations show strong evidence that we
live in a spatially flat universe \cite{Spergel:2003cb} with low
matter density \cite{Tegmark:2003ud}  that is currently undergoing
accelerated cosmic expansion
\cite{snobs,Riess:2004nr,Astier:2005qq}. The most direct
indication for the current accelerating expansion comes from the
accumulating type Ia supernovae (SnIa) data
\cite{Astier:2005qq,Riess:2004nr} which provide a detailed form of
the recent expansion history of the universe.

This accelerating expansion has been attributed to a dark energy
\cite{Sahni:2004ai} component with negative pressure which can
induce repulsive gravity and thus cause accelerated expansion.

The simplest and most obvious candidate for this dark energy is
the cosmological constant $\Lambda$ \cite{Sahni:1999gb} with
equation of state $w={p / \rho}=-1$. This model however raises
theoretical problems related to the fine tuned value required for
the cosmological constant \cite{Sahni:1999gb}. These difficulties
have lead to a large variety of proposed models where the dark
energy component evolves with time \cite{quintess} usually due to
an evolving scalar field (quintessence) which may be minimally
\cite{quintess} or non-minimally \cite{modgrav} coupled to
gravity. The main prediction of the dynamical models is the
evolution of the dark energy density parameter $\Omega_X(z)$.
Combining this prediction with the prior assumption for the matter
density parameter $\Omega_{0m}$, the predicted expansion history
$H(z)$ is obtained as \be H(z)^2 = H_0^2 [\Omega_{0m} (1+z)^3 +
\Omega_X (z)] \ee The dark energy density parameter is usually
expressed as \be \Omega_X (z)=\Omega_{0X} (1+z)^{3(1+w(z))} \ee
where $w(z)$ is related to $H(z)$ by
\cite{Nesseris:2004wj,Huterer:2000mj}  \be \label{wz3}
w(z)={{{2\over 3} (1+z) {{d \ln H}\over {dz}}-1} \over
{1-({{H_0}\over H})^2 \Omega_{0m} (1+z)^3}} \ee If the dark energy
can be described as an ideal fluid with conserved energy momentum
tensor $T^{\mu \nu}=diag(\rho, p, p, p) $ then the above parameter
$w(z)$ is identical with the equation of state parameter of dark
energy \be w(z)=\frac{p(z)}{\rho(z)} \ee Independently of its
physical origin, the parameter $w(z)$ is an observable derived
from $H(z)$ (with prior knowledge of $\Omega_{0m}$) and is usually
used to compare theoretical model predictions with observations.

Most evolution behaviors of $w(z)$ can be reproduced by assuming
appropriate scalar field quintessence potentials. If however
$w(z)$ were observationally found to cross the phantom divide line
(PDL) $w=-1$ then all  minimally coupled single scalar field
models would be ruled out as dark energy
candidates\cite{Vikman:2004dc,Perivolaropoulos:2004yr} (this
includes phantom\cite{phantom} and $k-essence$
models\cite{Armendariz-Picon:2000ah}). This would leave only
models based on extended gravity
theories\cite{Perivolaropoulos:2005yv,Tsujikawa} and combinations
of multiple fields \cite{Wei:2005nw,Anisimov:2005ne}(quintessence
+ phantom) as dark energy candidates. It is therefore important to
utilize the available SnIa datasets to place constraints on $w(z)$
and determine the likelihood of having a $w(z)$ that crosses the
PDL, 
as done in Ref.\cite{Huterer:2004ch} using the Gold dataset
following a robust procedure. This paper confirms the evidence for
crossing the PDL with $w<-1$ at present and also shows some hint
for oscillating $w(z)$ at best fit in agreement with
Ref.\cite{Nesseris:2004wj}, which were the the first to point this
trend.\enlargethispage{\baselineskip}
\enlargethispage{\baselineskip} \enlargethispage{\baselineskip}

The two most reliable and robust SnIa datasets existing at present
are the Gold dataset \cite{Riess:2004nr} and the Supernova Legacy
Survey (SNLS) \cite{Astier:2005qq} dataset. The Gold dataset
compiled by Riess et. al. is a set of supernova data from various
sources analyzed in a consistent and robust manner with reduced
calibration errors arising from systematics. It contains 143
points from previously published data plus 14 points with $z>1$
discovered recently with the HST. The SNLS is a 5-year survey of
SnIa with $z<1$. It has recently \cite{Astier:2005qq} released the
first year dataset. The SNLS has adopted a more efficient SnIa
search strategy involving a `rolling search' mode where a given
field is observed every third or fourth night using a single
imaging instrument thus reducing photometric systematic
uncertainties. The published first year SNLS dataset consists of
44 previously published nearby SnIa with $0.015<z<0.125$ plus 73
distant SnIa ($0.15<z<1$) discovered by SNLS two of which are
outliers and are not used in the analysis. 
The fact that in the two datasets a set of low-z SnIa is common to
both samples could only lead to minor common systematics due to
low redshift.

The above observations provide the apparent magnitude $m(z)$ of
the supernovae at peak brightness after implementing correction
for galactic extinction, K-correction and light curve
width-luminosity correction. The resulting apparent magnitude
$m(z)$ is related to the luminosity distance $d_L(z)$ through \be
m_{th}(z)={\bar M} (M,H_0) + 5 log_{10} (D_L (z)) \label{mdl} \ee
where in a flat cosmological model \be D_L (z)= (1+z) \int_0^z
dz'\frac{H_0}{H(z';a_1,...,a_n)} \label{dlth1} \ee is the Hubble
free luminosity distance ($H_0 d_L/c$), $a_1,...,a_n$ are
theoretical model parameters and ${\bar M}$ is the magnitude zero
point offset and depends on the absolute magnitude $M$ and on the
present Hubble parameter $H_0$ as \ba
{\bar M} &=& M + 5 log_{10}(\frac{c\; H_0^{-1}}{Mpc}) + 25= \nn \\
&=& M-5log_{10}h+42.38 \label{barm} \ea The parameter $M$ is the
absolute magnitude which is assumed to be constant after the above
mentioned corrections have been implemented in $m(z)$.

The data points of the Gold dataset are given after the
corrections have been implemented, in terms of the distance
modulus \be \mu^G_{obs}(z_i)\equiv m^G_{obs}(z_i) - M
\label{mug}\ee The SNLS dataset however also presents for each
point, the stretch factor $s$ used to calibrate the absolute
magnitude and the rest frame color parameter $c$ which mainly
measures host galaxy extinction by dust. Thus, the distance
modulus in this case depends apart from the absolute magnitude
$M$, on two additional parameters $\alpha$ and $\beta$ defined
from \be
\mu_{obs}^{SNLS}=m_{obs}^{SNLS}(z_i)-M+\alpha(s_i-1)-\beta c_i
\label{musnls} \ee which are fit along with the theoretical
parameters using a recursive procedure discussed below.

The theoretical model parameters are determined by minimizing the
quantity \be \chi^2 (a_1,...,a_n)= \sum_{i=1}^N
\frac{(\mu_{obs}(z_i) - \mu_{th}(z_i))^2}{\sigma_{\mu \; i}^2 +
\sigma_{int}^2 + \sigma_{v\; i}^2 } \label{chi2} \ee where
$\sigma_{\mu \; i}^2$, $\sigma_{int}^2$ and $\sigma_{v\; i}^2$ are
the errors due to flux uncertainties, intrinsic dispersion of SnIa
absolute magnitude and peculiar velocity dispersion respectively.
These errors are assumed to be gaussian and uncorrelated. The
theoretical distance modulus is defined as \be \mu_{th}(z_i)\equiv
m_{th}(z_i) - M =5 log_{10} (D_L (z)) +\mu_0 \label{mth} \ee where
\be \mu_0= 42.38 - 5 log_{10}h \label{mu0}\ee and $\mu_{obs}$ is
given by (\ref{mug}) and (\ref{musnls}) for the Gold and SNLS
datasets respectively.

The steps we followed for the minimization of (\ref{chi2}) for the
Gold and SNLS datasets are described in detail in the Appendix.
The validity of our analysis has been verified by comparing the
part of our results that overlaps with the results of the original
Refs \cite{Astier:2005qq,Riess:2004nr} of the Legacy and Gold
datasets.

\section{Comparative Analysis}

We will consider four representative $H(z)$ parametrizations and
minimize the $\chi^2$ of eq. (\ref{chi2}) with respect to model
parameters. We compare the best fit parametrizations obtained with
three datasets
\begin{itemize} \item The full SNLS dataset with 115 datapoints
(excluding two outliers) and $z<1$.

\begin{figure*}
\begin{center}
\rotatebox{270}{\resizebox{.8\textwidth}{!}{\includegraphics{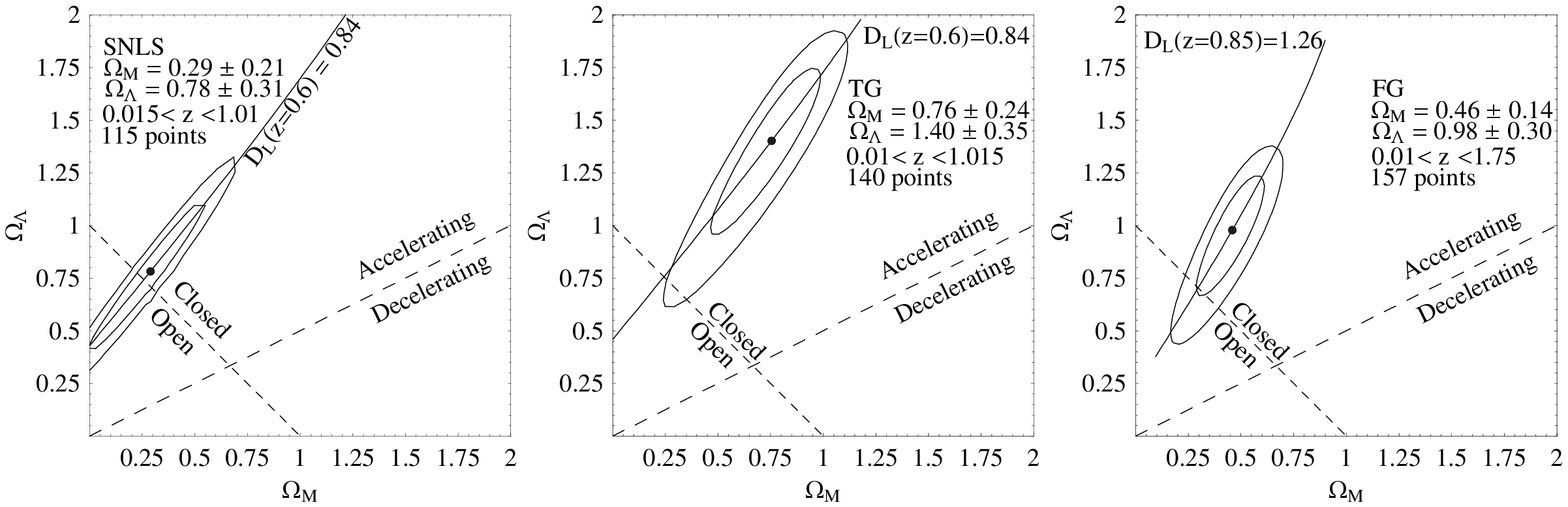}}}
\vspace{-250pt}{ \caption{The $ 68 \%$ and $ 95 \% $ confidence
region ellipses in the $\Omega_{0m}-\Omega_{\Lambda}$ plane based
on parametrization (\ref{lcdm}). The three plots correspond to the
three datasets discussed in the text (SNLS, TG and FG). Notice
that flat LCDM is more favored by the SNLS dataset than by the
Gold datasets.}} \label{fig1}
\end{center}
\end{figure*}

\item The Full Gold
dataset (FG) with 157 datapoints and $0<z<1.7$. \item A Truncated
version of the Gold dataset (TG) with 140 datapoints and $z<1$
which can be compared in a more direct way with SNLS.
\end{itemize} The four fitted parametrizations include the general
LCDM without a flat prior: \ba H(z)^2 &=& H_0^2 [\Omega_{0m}
(1+z)^3 + \Omega_\Lambda + \nn \label{lcdm} \\ & +&(1- \Omega_{0m}-\Omega_\Lambda)(1+z)^2]
\\ w(z)&=&-1 \ea and three dynamical dark energy
parametrizations with two free parameters which allow for crossing
of the PDL:
\begin{itemize}
\item Parametrization A: \ba w(z)&=& w_0 +w_1 \; z \label{a1} \\
H^2 (z)&=&H_0^2 [\Omega_{0m} (1+z)^3+\\&+& \nn
(1-\Omega_{0m})(1+z)^{3(1+w_0-w_1)}e^{3w_1 z}]\ea \item
Parametrization B: \ba  w(z)&=&w_0+w_1 \frac{z}{1+z} \label{lda}
\\ H^2 (z)&=&H_0^2 [ \Omega_{0m} (1+z)^3
+ \nn \\ +(1-\Omega_{0m}) (1&+&z)^{3(1+w_0+w_1)}e^{3w_1[1/(1+z)-1]}]\ea
\item Parametrization C: \ba w(z)&=& \frac{a_1+3(\Omega_{0m}-1)-2a_1 z-a_2(-2+2z+z^2)}{3(1-\Omega_{0m}+a_1 z+2a_2
 z+a_2z^2)} \\
H^2 (z)&=&H_0^2 [\Omega_{0m} (1+z)^3 +
a_1(1+z)+a_2(1+z)^2+\;\;\;\;\;\;\;\; \nn \\
&+&(1-\Omega_{0m}-a_1-a_2)] \label{c2}\ea
\end{itemize}

The motivation behind parametrization C is to mimic a
two-component DE model. Alternatively, it could be viewed as a
power law expansion in the scale factor dependence of the DE
energy density. In analyzing
the general LCDM of eq.(\ref{lcdm}) we used \ba & & D_L(z)=  \\
&=& \frac{(1 + z)}{\sqrt{\Omega_{0m}+\Omega_{0X}-1}} Sin
[\sqrt{\Omega_{0m}+\Omega_{0X}-1} \int_0^z dz \frac{H_0}{H(z)}]\nn
\label{nfdl} \ea instead of eq.(\ref{dlth1}) which is only
suitable for flat models.

In Fig. 1 we show the confidence region ellipses in the
$\Omega_{0m}-\Omega_{\Lambda}$ plane based on parametrization
(\ref{lcdm}). The three plots correspond to the three datasets
discussed above (SNLS, TG and FG). The following comments can be
made on these plots:
\begin{itemize}
\item The major axes of the elliptic contours are approximately
parallel in the three plots. This effect\cite{Choudhury:2003tj} is
due to the degeneracy of the fitted
$D_L(z;\Omega_{0m},\Omega_\Lambda)$ with respect to certain linear
combinations of the parameters $\Omega_{0m}-\Omega_{\Lambda}$ in
the redshift range of interest. For example, choosing a
representative redshift $z=0.6$ it is easy to show that the value
$D_L (z=0.6)=0.84$ is obtained by all combinations of
$\Omega_{0m}-\Omega_{\Lambda}$ that satisfy $\Omega_{0m} -0.80
\Omega_\Lambda=-0.38$. The direction of this (approximate)
degeneracy line is determined by the $H(z)$ parametrization and
the redshift range considered but the actual location of the line
is determined by the data. \item The two versions of the Gold
dataset favor a closed universe instead of a flat universe
($\Omega_{tot}^{TG}=2.16\pm 0.59$, $\Omega_{tot}^{FG}=1.44\pm
0.44$). This trend is not realized by the SNLS dataset which gives
$\Omega_{tot}^{SNLS}=1.07\pm 0.52$. \item The point corresponding
to SCDM $(\Omega_{0m},\Omega_\Lambda)=(1,0)$ is ruled out by all
datasets at a confidence level more than $10\sigma$.

\begin{figure*}
\begin{center}
\rotatebox{270}{\resizebox{0.70\textwidth}{!}{\includegraphics{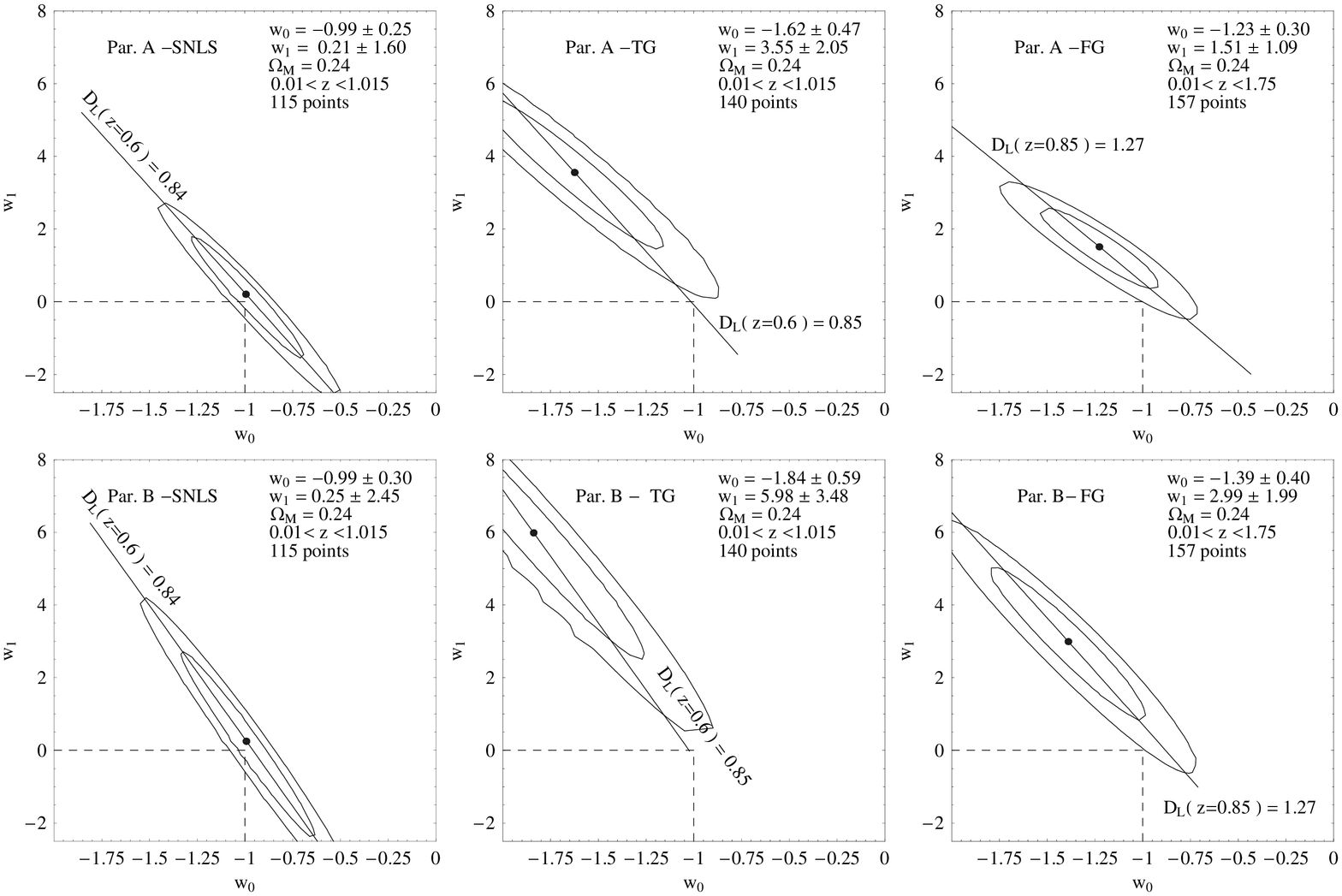}}}
\caption{The $68\%$ and $95\%$ $\chi^2$ confidence contours of
parametrizations A and B assuming a flat prior using the datasets
SNLS, TG and FG. A prior of $\Omega_{0m}=0.24$ has been used. The
dashed lines intersect at the parameter values of flat LCDM.
Notice that for the SNLS, flat LCDM almost coincides with the best
fit for both parametrizations.} \label{fig2}
\end{center}
\end{figure*}

\item The values of $\mu_0^{min}$ that minimize the
$\chi^2(\Omega_{0m},\Omega_\Lambda)$ of eq. (\ref{chi2})  (with
$H(z)$ given by (\ref{lcdm})) obtained by all three datasets are
consistent with each other. We find $\mu_0^{SNLS}=43.15\pm 0.05$,
$\mu_0^{TG}=43.30\pm 0.05$ and $\mu_0^{FG}=43.32\pm 0.05$. This
alleviates the possible discrepancy between high and low redshift
data discussed recently in Ref. \cite{Choudhury:2003tj}. \item If
we use a prior constraint of flatness
$\Omega_{0m}+\Omega_\Lambda=1$ thus restricting on the
corresponding dotted line of Fig. 1 and using the parametrization
\be H(z)^2=H_0^2 [\Omega_{0m} (1+z)^2 + (1-\Omega_{0m})] \ee we
find minimizing $\chi^2(\Omega_{0m})$ of eq (\ref{chi2}) \ba
\Omega_{0m}^{SNLS} &=& 0.26\pm 0.04 \\
\Omega_{0m}^{TG} &=& 0.30\pm 0.05 \\
\Omega_{0m}^{FG} &=& 0.31\pm 0.04 \ea

The values of $\Omega_{0m}^{SNLS}$ and $\Omega_{0m}^{FG}$ are
practically identical with the corresponding in the original Refs
\cite{Riess:2004nr,Astier:2005qq} where the data were first
published.

\begin{figure*}
\begin{center}
\rotatebox{270}{\resizebox{0.70\textwidth}{!}{\includegraphics{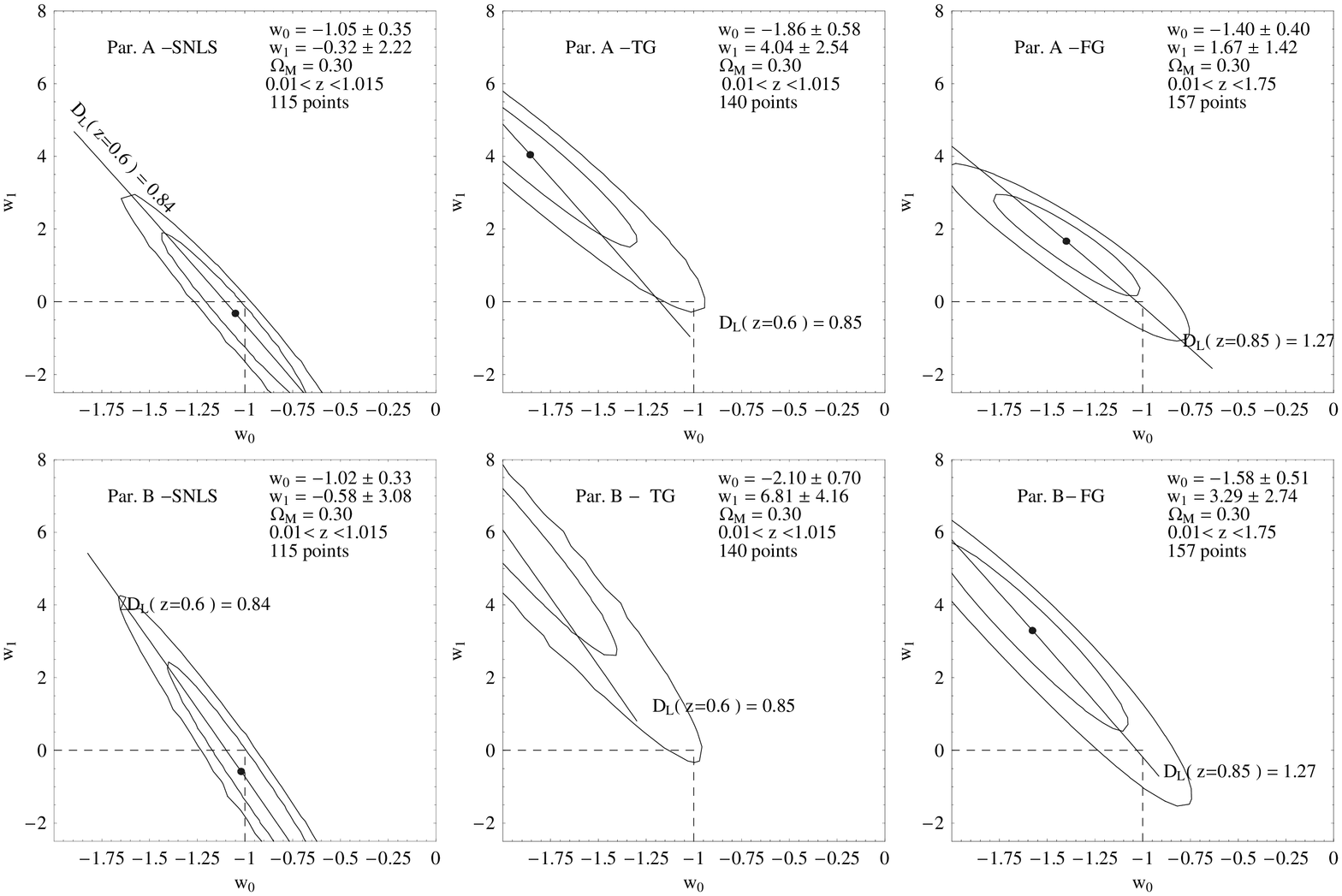}}}
\caption{Same as Fig. 2 but with a prior $\Omega_{0m}=0.3$.}
\label{fig3}
\end{center}
\end{figure*}

This, along other similar tests, confirms the validity of our
analysis.\end{itemize}

Even though the cosmological constant with equation of state
parameter $w=-1$ is the simplest form of dark energy consistent
with the data, the possibility of evolving dark energy models with
non-constant $w(z)$ remains a viable alternative which may even
provide better fits to the data than LCDM. To address this issue
we considered the three parametrizations (A, B, C) of eqs
(\ref{a1})-(\ref{c2}) and assuming flatness we constrained their
parameters using the three datasets.

Parametrizations A and B reduce to $w(z)=w_0$ in the special case
when $w$ does not evolve with time ($w_1=0$).

With this prior we construct Table I showing the best fit $w_0$
value and $1\sigma$ errors for each dataset and
$\Omega_{0m}=0.24-0.3$. Clearly all datasets are consistent with
each other ruling out models with $w>1/3$ at more than a
$10\sigma$ confidence level.

\vspace{0pt}
\begin{table}[b!]
\begin{center}
\caption{The best fit $w_0$ parameter values for each dataset
assuming priors of flatness and $w_1=0$. \label{table1}}
\begin{tabular}{cccc}
\hline
\hline\\
\bf{Matter Density  $\Omega_{0m}$}&\bf{SNLS}& \bf{TG}& \bf{FG}\\
\\\hline \vspace{0pt}\\
0.24 & $w_0=-0.95\pm 0.09$    \hspace{7pt}& $w_0=-0.89\pm 0.10$  \hspace{7pt} & $w_0=-0.86\pm 0.09$ \hspace{7pt} \\
0.30 & $w_0=-1.11\pm 0.11$   \hspace{7pt}& $w_0=-1.04\pm 0.12$  \hspace{7pt} & $w_0=-1.02\pm 0.11$\hspace{7pt} \\
 \hline \hline
\end{tabular}
\end{center}
\end{table}

Extending the analysis to the full parameter space we construct
$\chi^2$ confidence contours in Figs. 2 and 3 assuming a flat
prior.

In order to investigate the dependence of our results on the prior
of $\Omega_{0m}$ we do not marginalize over it. Instead we present
the $w_0-w_1$ confidence contours of parametrizations A and B for
the priors $\Omega_{0m}=0.24$ (Fig. 2) and $\Omega_{0m}=0.3$ (Fig.
3).  The point corresponding to LCDM is also shown in these
figures. For parametrization C we have not presented the
confidence contours because $H(z)$ becomes complex for regions of
parameter space overlaping with the 95\% confidence contours.

The following comments can be made regarding Figs. 2 and 3.
\begin{itemize}
\item The minimal LCDM model ($w_0=-1$, $w_1=0$) appears to be
close to the 95\% confidence contour in the analyses based on the
TG and FG datasets.  For the analysis based on the SNLS however,
the flat LCDM model is well within the 68\% contour and in fact
for $\Omega_{0m}=0.24$ it is almost identical with the best fit
parametrization in both the A and B parametrization cases! Thus
LCDM appears to have significantly gained in likelihood compared
to dynamical dark energy models in the context of the new SNLS
dataset.\end{itemize}
\newpage

\begin{figure*}[t!]
\begin{center}
\rotatebox{270}{\resizebox{0.95\textwidth}{!}{\includegraphics{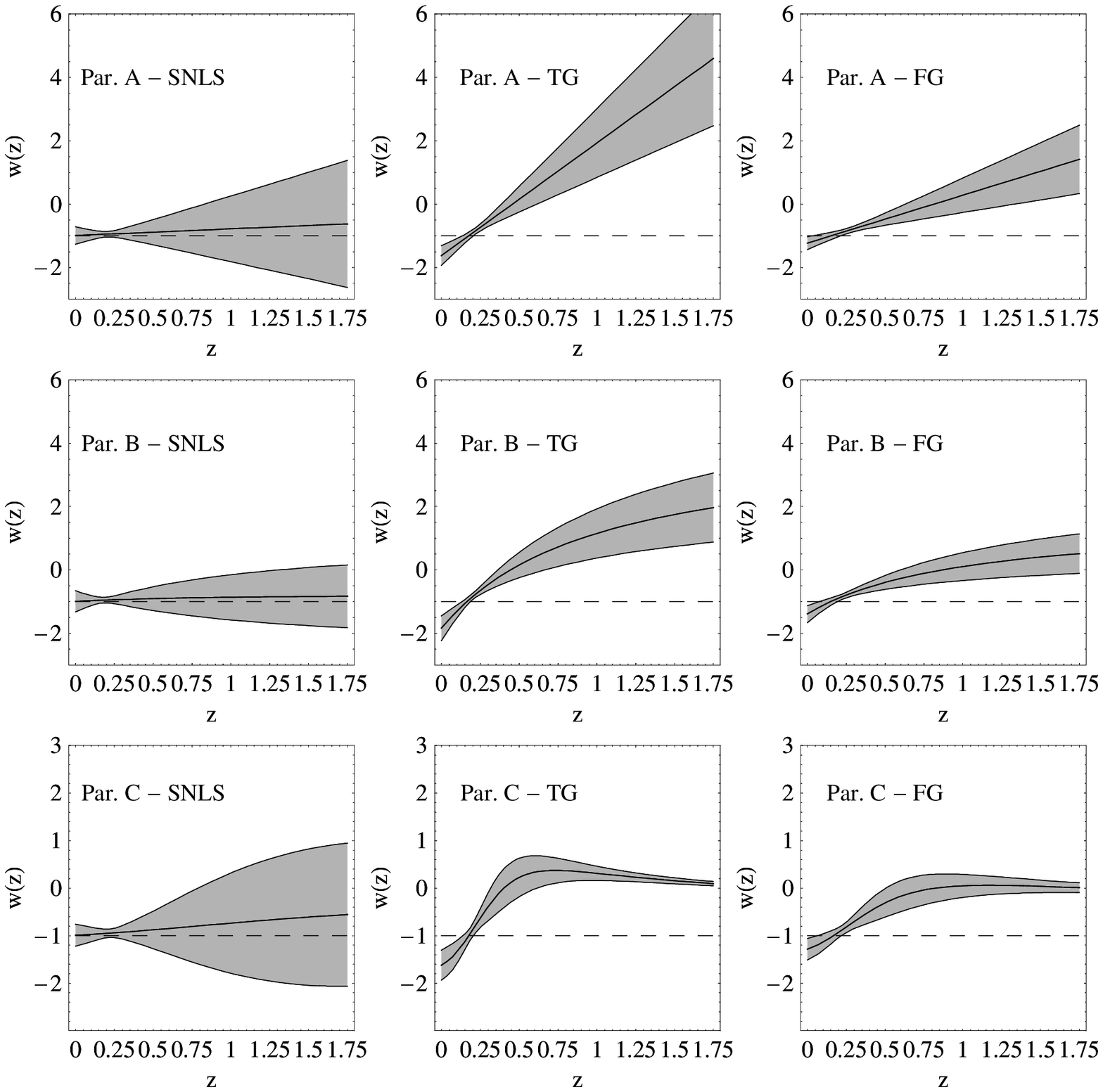}}}
\caption{The best fit $w(z)$ are plotted for each dataset (SNLS,
TG and FG) and each parametrization (A, B and C) with a prior of
$\Omega_{0m}=0.24$. The thick solid line represents the best-fit
and the light grey contour represents the $1 \sigma$ confidence
level around the best-fit. The dased horizontal line represents
LCDM. Notice that PDL crossing at best fit occurs only for the TG
and the FG (Gold) datasets while the best fits of the SNLS dataset
almost coincide with LCDM.} \label{fig4}
\end{center}
\end{figure*}

\begin{itemize}
\item The best fit $w(z)$ parametrizations in the context of the
FG and the TG datasets, not only are far from the LCDM point
($w_0=-1$, $w_1=0$) (see Figs. 2 and 3) but they also clearly
cross the PDL line. This is demonstrated in Fig. 4 where the best
fit $w(z)$ are plotted for each dataset and each parametrization
with a prior of $\Omega_{0m}=0.24$.

This crossing of the PDL is not realized for the best fit A, B and
C parametrizations in the context of the SNLS dataset. Several
authors\cite{Wei:2005nw,Perivolaropoulos:2005yv,Hu:2004kh} have
been recently motivated by the high likelihood of the PDL crossing
indicated by the Gold datasets\cite{Alam:2004jy,phant-obs2} to
explore theoretical models that predict such crossing. It has been
shown that this task is not trivial and can not be achieved by a
single minimally coupled field\cite{Vikman:2004dc}.

Here we show however that such PDL crossing is not favored by the
new SNLS dataset and therefore the motivation for the above papers
is weakened. Indeed, phantom\cite{phantom} dynamical dark energy
models with $w_0 < -1$ are not favored by the SNLS dataset in
contrast with the Gold dataset that favored such models (see Figs
2 and 3 and Ref. \cite{phant-obs2,Lazkoz:2005sp}). \item Even
though the best fit parameter values are relatively insensitive to
the value of $\Omega_{0m}$ in the range $0.2<\Omega_{0m}<0.3$ the
errors are more sensitive increasing with $\Omega_{0m}$. Our
results for the best fit parameters of the parametrizations A, B
and C are shown in the legends of Figs 2 and 3. For both
$\Omega_{0m}$ priors  the errors of $w_1$ are much larger compared
to those of $w_0$. As discussed above, this is a generic feature
of the parametrization used and is not related to the particular
datasets.
\end{itemize}

\section{Discussion-Conclusion}

We have performed a comparative analysis of the three most recent
and reliable SnIa datasets available in the literature: the Full
Gold (FG) dataset (157 data points $0<z<1.7$), the Truncated Gold
(TG) dataset (140 data points $0<z<1$) and the most recent SNLS
dataset (115 data points $0<z<1$). Our analysis is an extension of
our earlier analyses which had focused \cite{Lazkoz:2005sp} on the
FG and earlier \cite{Nesseris:2004wj} datasets. We have used
representative dark energy parametrizations to examine the
consistency among the three datasets in constraining the
corresponding parameter values. We have found that even though the
constraints obtained using the three datasets are consistent with
each other at the 95\% confidence level, the latest (SNLS) dataset
shows distinct trends which are not shared by the other (earlier)
datasets. The most characteristic of these trends are the
following:

\begin{itemize}
\item The best fit dynamical $w(z)$ obtained from the SNLS dataset
does not cross the PDL $w=-1$ and remains close to the $w=-1$ line
for the whole redshift range $0<z<1$. The LCDM parameters
($w_0=-1$, $w_1=0$) almost coincide with the best fit parameters
of the dynamical $w(z)$ parametrizations. Thus, the data do not
seem to require and utilize the additional dynamical parameters
offered to them. This is an interesting new feature of the data
which favors the minimal LCDM model. In contrast, the best fit
dynamical $w(z)$ obtained from the Gold datasets (FG and TG)
clearly crosses the PDL and departs significantly from the PDL
$w=-1$ line (see Fig. 4). According to these datasets the minimal
LCDM is consistent but is not favored. It is about $2\sigma$ away
from the best fit $w(z)$ which crosses the PDL.

\item The best fit $(\Omega_{0m},\Omega_\Lambda)$ parameters in a
LCDM parametrization without a flat prior show interesting
differences between the Gold and the SNLS datasets. In particular,
the SNLS favors a flat universe much more than the Gold datasets.
However, all three datasets remain consistent with flat LCDM at
the 95\% confidence level while SCDM is excluded by all datasets
at more than $10\sigma$.
\end{itemize}
The above mild trend differences between the Gold and the SNLS
datasets can be summarized by stating that the SNLS hints towards
the minimal flat LCDM concordance model more than the Gold
datasets. It is an exciting prospect to see whether this trend
will continue and get verified by upcoming future SnIa
observations.

\section{Appendix}
Here we give some details of our analysis for both the Gold and
the Legacy datasets. The full numerical analysis was performed
using Mathematica and it is available at
http://leandros.physics.uoi.gr/snls.htm

We first describe the method used for the Gold dataset analysis.
From eqs. (\ref{chi2}), (\ref{mth}) and (\ref{mu0}) we find \be
\chi^2 (a_1,...,a_n)=\sum_{i=1}^N \frac{(\mu_{obs\; i}-5
log_{10}D_L(z_i;a_1,...,a_n)-\mu_0)^2}{\sigma_i^2}
\label{chi2g}\ee where \be \sigma_i^2=\sigma_{\mu \; i}^2 +
\sigma_{int}^2 + \sigma_{v\; i}^2 \ee is the total error published
for the Gold dataset.

The parameter $\mu_0$ is a nuisance parameter but it is
independent of the data points and the dataset. This expected
independence can be used as consistency test of the data
\cite{Choudhury:2003tj}. The minimization with respect to $\mu_0$
can be made trivially by expanding  the $\chi^2$ of equation
(\ref{chi2}) with respect to $\mu_0$ as \be \chi^2 (a_1,..,a_n) =
A - 2 \mu_0 B  + \mu_0^2 C \label{chi2bm} \ee where \ba
A(a_1,..,a_n)&=&\sum_{i=1}^{N} \frac{(m_{obs}(z_i) -
m_{th}(z_i ;\mu_0=0,a_1,..,a_n))^2}{\sigma_{m_{obs}(z_i)}^2} \label{bb} \nn \\
B(a_1,..,a_n)&=&\sum_{i=1}^{N} \frac{(m_{obs}(z_i) - m_{th}(z_i
;\mu_0=0,a_1,..,a_n))}{\sigma_{m_{obs}(z_i)}^2} \label{bb} \nn \\
C&=&\sum_{i=1}^{N}\frac{1}{\sigma_{m_{obs}(z_i)}^2 } \label{cc}
\ea Equation (\ref{chi2bm}) has a minimum for $\mu_0={B}/{C}$ at
\be {\tilde\chi}^2(a_1,...,a_n)=A(a_1,...,a_n)-
\frac{B(a_1,...,a_n)^2}{C} \label{chi2min1}\ee Thus instead of
minimizing $\chi^2(\mu_0,a_1,...,a_n)$ we can minimize
${\tilde\chi}^2(a_1,...,a_n)$ which is independent of $\mu_0$.
Obviously $\chi_{min}^2={\tilde\chi}_{min}^2$ and this is the
approach used. 
Alternatively we could have performed a uniform marginalization
over the nuisance parameter $\mu_0$ thus obtaining
\cite{Nesseris:2004wj,Perivolaropoulos:2004yr,DiPietro:2002cz} \be
{\tilde\chi}^2(a_1,...,a_n)=A(a_1,...,a_n)-
\frac{B(a_1,...,a_n)^2}{C} + \ln (C/2\pi) \label{chi2min2}\ee to
be minimized with respect to $a_1,...,a_n$. In our Gold dataset
analysis we consider the ${\tilde\chi}^2(a_1,...,a_n)$ of equation
(\ref{chi2min1}) which is already minimized with respect to
$\mu_0$. 
If we marginalized over all values of $H_0$, as in Ref.
\cite{Riess:2004nr}, that would just add a constant (see
eq.(\ref{chi2min2})) and would not change the results. The
minimization of (\ref{chi2g}) was made using the FindMinimum
command of Mathematica.

Our analysis of the SNLS dataset proceeded in a somewhat different
manner following Ref. \cite{Astier:2005qq}. Using equation
(\ref{musnls}) in (\ref{chi2}) we constructed $\chi^2$ as \ba
\chi^2 (\alpha,\beta,M+\mu_0, a_1,...,a_n)=\\ \nn =\sum_{i=1}^N
\frac{(\mu_{obs\; i}-5
log_{10}D_L(z_i;a_1,...,a_n)-\mu_0)^2}{\sigma_{\mu \; i}^2 +
\sigma_{int}^2 + \sigma_{v\; i}^2} \label{chi2snls} \ea For the
distance modulus error we have \be \sigma_{\mu \; i}^2=\sigma_{m
\; i}^2 + \alpha^2 \sigma_{s \; i}^2 + \beta^2 \sigma_{c \; i}^2
\ee Each one of the $\sigma_{m \; i}$, $\sigma_{s \; i}$ and
$\sigma_{c \; i}$ has been published in Ref. \cite{Astier:2005qq}.
The velocity dispersion error $\sigma_v^2$ assuming a peculiar
velocity dispersion of $300 km/sec$ may be written as \be
\sigma_{v\;i}^2=\frac{5\;10^{-3}}{ln(10)}(\frac{1}{1+z_i}+\frac{1}{H(z_i)\int_0^{z_i}\frac{dz}{H(z)}})^2
\ee The intrinsic dispersion error $\sigma_{int}^2$ is initially
set to a value $\sigma_{int}=0.15$ and then updated with the
following three step procedure \cite{Astier:2005qq}:
\begin{enumerate}
\item Fix the values of $\alpha$ and $\beta$ in $\sigma_{\mu \;
i}^2$ and minimize the $\chi^2$ of eq. (\ref{chi2snls}) with
$\sigma_{int}=0.15$. If this fixing is not made, a bias is
introduced towards increasing errors during minimization. \item
Change the value of $\sigma_{int}$ to obtain $\chi^2 =1$. \item
Use the new value of $\sigma_{int}$ and minimize again keeping the
values of $\alpha$ and $\beta$ in $\sigma_{\mu \; i}^2$ fixed.
\end{enumerate}
This procedure, with no marginalization over $M+\mu_0$, $\alpha$,
$\beta$ as described in page 10 of Ref. \cite{Astier:2005qq},
leads to the best fit values of $M+\mu_0$, $\alpha$, $\beta$,
$a_1,...,a_n$.

The errors are evaluated using the covariance
matrix of the fitted parameters \cite{press92} 
and the errors on the equation of state $w(z;p_i)$ are given by
\be \sigma_w^2 =\sum_{i=1}^n(\frac{\partial w}{\partial
p_i})C_{ii}+ 2\sum_{i=1}^n\sum_{j=i+1}^n (\frac{\partial
w}{\partial p_i}) (\frac{\partial w}{\partial p_j}) C_{ij} \ee
where $p_i$ are the cosmological parameters and $C_{ij} $ the
covariance matrix \cite{Alam:2004ip}.

\section*{Acknowledgements}
 This work was supported by the program
PYTHAGORAS-1 of the Operational Program for Education and Initial
Vocational Training of the Hellenic Ministry of Education under
the Community Support Framework and the European Social Fund. SN
acknowledges support from the Greek State Scholarships Foundation.

\end{document}